\documentclass{mem}
\usepackage{natbib}\usepackage{txfonts}\usepackage{balance}
\usepackage{graphicx}
\usepackage[a4paper]{hyperref}
\idline{75}{282}
\begin{document}
\def\teff{$T\rm_{eff }$}
\def\kms{$\mathrm {km s}^{-1}$}
\def\msun{\rm M_{\odot}}
\def\etal{{et al.\ }}
\def\simlt{\mathrel{\rlap{\lower 3pt\hbox{$\sim$}}\raise 2.0pt\hbox{$<$}}}
\def\simgt{\mathrel{\rlap{\lower 3pt\hbox{$\sim$}} \raise 2.0pt\hbox{$>$}}}
\def\lsim{\mathrel{\rlap{\lower 3pt\hbox{$\sim$}}\raise 2.0pt\hbox{$<$}}}
\def\gsim{\mathrel{\rlap{\lower 3pt\hbox{$\sim$}} \raise 2.0pt\hbox{$>$}}}
\def\di{\mbox{d}}
\def\mbulge{M_{\rm Bulge}}
\def\msunpc3{\msun~{\rm {pc^{-3}}}}
\newcommand{\be}{\begin{equation}}
\newcommand{\ee}{\end{equation}}

\title{Massive black hole binaries in gaseous nuclear discs}

   \subtitle{}

\author{
M. \,Dotti\inst{1}, M. \,Colpi\inst{2}, F. \,Haardt\inst{3}, L. Mayer\inst{4,}\inst{5} 
}

\institute{
Department of Astronomy, University of Michigan, Ann Arbor, MI 48109, USA
\email{mdotti@umich.edu}
\and
Dipartimento di Fisica G.~Occhialini, Universit\`a degli Studi di Milano
Bicocca, Piazza della Scienza 3, 20126 Milano, Italy
\and
Dipartimento di Fisica e Matematica, Universit\`a dell'Insubria, Via Valleggio 11, 22100 Como, Italy
\and
Institute for Theoretical Physics, University of Zurich, CH-8057, Zurich, Switzerland
\and
Institute of Astronomy, Department of Physics, ETH, Zurich, Wolfgang-Pauli Strasse, CH-8095 Zurich, Switzerland
}

\authorrunning{Dotti}

\titlerunning{Massive black hole binaries in nuclear discs}

\abstract{ We  study the evolution of  a massive black hole  pair in a
rotationally supported  nuclear disc.  The distributions of  stars and
gas mimic  the nuclear  region of a  gas--rich galaxy  merger remnant.
Using  high--resolution  SPH simulations,  we  follow  the black  hole
dynamics and trace the evolution of the underlying background, 
until the black holes form a binary. We find  that the
gravitational perturbation of the pair creates 
a core in the disc density profile, hence decreasing the gas-dynamical drag. 
This leads the newly formed binary to stall at a separation of $\sim \,5$  pc.  In the early
phases of the sinking, black  holes   lose  memory  of  their  initial  orbital
eccentricity  if  they  co--rotate   with  the  disc, as rotation of the gaseous background 
promotes circularization of the black hole orbits. Circularization is
efficient until the black holes bind in a binary, though in the latest stages
of  the simulations  a residual  eccentricity $\gsim\,  0.1$  is still
present. Black holes are treated as sink particles, 
allowing for gas accretion. 
We find  that accretion strongly depends on the dynamical
properties  of the black  holes,  and occurs preferentially after circularization. 
}
\maketitle{}

\section{Introduction}

Collisions of gas rich spiral galaxies may trigger starbursts as those 
observed in luminous infrared galaxies (LIRGs). 
A large number of LIRGs hosts a central rotationally supported
massive (up to $10^{10}\msun$) gaseous disc extending on scales of
$\sim 100$ pc (Sanders \& Mirabel 1996; Downes \& Solomon 1998). 
These discs may be the end--product of 
gas--dynamical, gravitational torques excited during the merger, when 
large amounts of gas is driven into the core of the remnant  
(Kazantzidis et al. 2005; Mayer
et al. 2007).
 
Inside a massive self--gravitating disc, a putative MBH pair can continue
its dynamical evolution, and, possibly, can accrete gas, producing an observable
double AGN (Kocsis et al. 2005, Dotti et al. 2006).
Here we study the role of a nuclear gaseous disc
in driving the orbital evolution of the MBH binary, and asses 
the possibility of gas accretion into each pair member.

\section{Simulation setup}

We follow the dynamics of MBH pairs in nuclear discs using numerical
simulations run with the N--Body/SPH code GADGET (Springel, Yoshida \&
White 2001).

In our models, two MBHs are placed in the plane of
a  gaseous  disc, embedded  in  a  larger  scale stellar  spheroid.
The gaseous disc is modeled with 2 $\times 10^6$ particles, has a
total mass $M_{\rm{Disc}}=10^8 \msun$, and follows a Mestel surface
density profile. 
The disc is
rotationally supported, and has a radial and vertical scale of 100
and 10 pc, respectively. 
SPH particles evolve adiabatically (with a polytrophic index = 5/3),
and we set the initial internal energy density profile so that the
Toomre parameter of the disc
is $> 3$ everywhere, preventing disc fragmentation and formation of
large scale over--densities, such as bars and spiral arms. 

The spheroidal component (bulge) is modeled with $10^5$ collisionless
particles, initially distributed as a Plummer sphere, with a core radius of $50$ pc, and 
a total mass $M_{\rm Bulge}=6.98M_{\rm Disc}$. 

The two MBHs are equal mass ($M_{\rm BH}=4\times 10^6\,\msun$). 
$M_1$ is placed at rest at the centre of the circumnuclear disc, while $M_2$ 
is initially orbiting in the plane of the disc on an orbit whose eccentricity is $\simeq 0.7$, 
at a separation of 50 pc from $M_1$. $M_2$ can either be co-- or counter--rotating with respect to the
circumnuclear disc (runs A and B, respectively). 

We allow the gas particles
to be accreted onto the MBHs if the two following criteria are fulfilled:\\
$\bullet$ the total energy (kinetic + internal + gravitational) of the gas
particle is lower than 7/10 of its gravitational energy (all the energies
are computed with respect to each MBH)\\
$\bullet$ the total mass accreted onto a MBH every timestep is lower than the
$\dot{M}$ corresponding to an Eddington luminosity ($L_{\rm Edd}$)
assuming a radiative efficiency of 10\%. 

The spatial resolution of the
hydrodynamical force in the highest density regions is $\approx 0.1$ pc. We
set the gravitational softening for the gaseous particles and the MBHs 
at the same value to 
prevent numerical errors due to the different resolution. 
With this spatial
resolution we can resolve the influence radius of $M_1$ ($\approx 1$ pc),
a condition necessary to asses gas accretion 
\footnote{The influence radius of $M_2$ depends on the phase of its orbit.}.

\section{MBHs dynamical evolution} 

\begin{figure}[]
\resizebox{\hsize}{!}{\includegraphics[clip=true]{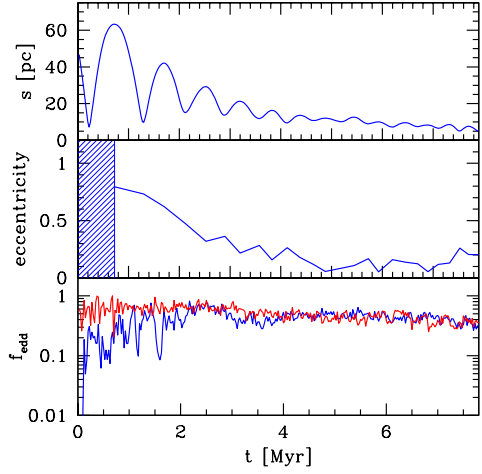}}
\caption{
\footnotesize
Run A. Upper panel: MBH separation as a function of time.
Middle panel: MBH pair eccentricity evolution. The shaded area refers to
the time before the first apocentre, when we can not evaluate the
value of the eccentricity. Lower panel: Eddington accretion ratio as
a function of time. Red and blue lines refer to M1 and M2, respectively. 
}
\label{dyn1}
\end{figure}

The upper panel of Fig.~\ref{dyn1} shows the separation
between the two MBHs as a function of time for run A.
The two MBHs reach a separation of the order of few pc in less than 10
Myr. After an initial fast migration, the shrinking process becomes inefficient and the
binary stalls at a separation of $\approx 5$ pc, larger than our force resolution. 

The eccentricity evolution of the MBH pair is presented in the middle
panel of Fig.~\ref{dyn1}. The MBH pair loses memory of its initial eccentricity
in the early phases of the orbital decay, when the two MBH are at a 
separation of $\sim 10$ pc. Such circularization happens because of 
the different effects that dynamical friction exerts on $M_2$ at different orbital phases 
(see  Dotti, Colpi \& Haardt 2006; Dotti et al. 2007 for a detailed review).
Dynamical friction is efficient in
reducing the eccentricity down to values $<0.1$,  while,  when the binary
forms, eccentricity grows again, up to $\approx 0.1$. 

\begin{figure}[]
\resizebox{\hsize}{!}{\includegraphics[clip=true]{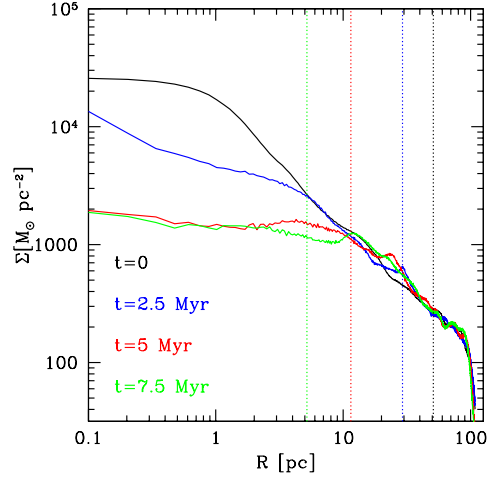}}
\caption{ \footnotesize Surface
density profile of the circumnuclear disc at four different times in
run A.  The solid black, blue, red and green solid lines refer to $t=0$, 2.5,
5, and 7.5 Myr respectively. The vertical dotted lines refer to the current
position of $M_2$ at the four times.  
}
\label{dyn2}
\end{figure}

The stalling of the MBH pair and the residual eccentricity are due to the
formation of a central core in the gaseous disc, as shown in
Fig.~\ref{dyn2}. The primary cause of the core formation is energy and angular momentum
transfer from the orbiting MBH to the disc due to the dynamical friction. 
The inner core forms during the early phases of the MBH orbital
decay, when $M_2$ orbit is still eccentric. We note that the interaction between the MBHs and
the circumnuclear disc in this phase was not fully resolved
in Dotti et al. (2007). As a check of our initial conditions, we evolved the disc
without $M_2$ for $\sim 10^7$ yr, and no core formation was observed. 
We also ran the same simulation with accretion switched off, and we found no differences in  
the dynamical evolution of the MBH pair, as expected due to the small amount of gas accreted by the MBHs with
respect to the total mass of the circumnuclear disc ($\sim 0.1\%$). 

For the counter-rotating case (Run B)
we find that the two MBHs also stall at a separation of
$\approx 5$ pc with a non zero eccentricity as illustrated in 
Fig.~\ref{acc2}.  The
middle panel of Fig.~\ref{acc2} shows the evolution of the
z--component of the orbital angular momentum  $L_{\rm z}$ of $M_2,$ 
normalized to its initial value. The angular momentum (initially negative) 
grows very efficiently during the first Myr, when $M_2$
is passing through the central, high density region of the disc. 
Angular momentum continues to
grow monotonically for the next $3-4$ Myrs, then becomes positive
($L_{\rm z}\,\approx 0.1$). That is, $M_2$ starts to
move on a co--rotating orbit with respect to the disc. The dynamical
friction process is the ultimate cause of this orbital ``angular momentum flip''.

\section{Accretion processes} 

\begin{figure}[]
\resizebox{\hsize}{!}{\includegraphics[clip=true]{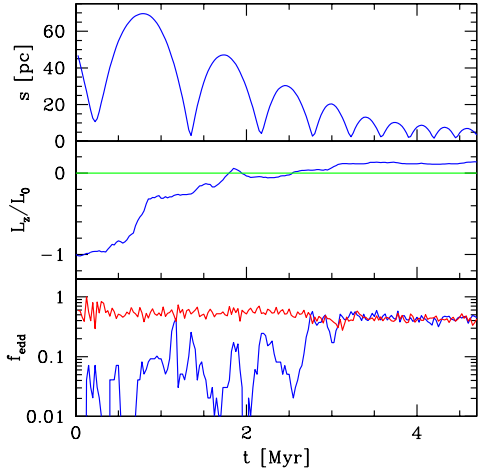}}
\caption{ \footnotesize 
Run B. Upper panel: MBHs separation as a function of time.
Lower panel: Eddington accretion ratio as a function of time. Red and blue
lines refer to $M_1$ and $M_2$, respectively.
}
\label{acc2}
\end{figure}

Fig.~\ref{dyn1} and \ref{acc2} allow to compare directly
the dynamical properties of the two MBHs to the
accretion rates in runs A and B respectively. 
The upper panels show the MBHs separation, while the lower
panels show the Eddington ratio 
$f_{\rm Edd}\equiv \dot{M} / \dot{M}_{\rm Edd}$ 
for the central (red lines) and orbiting (blue lines) MBHs. 
In both runs, $M_1$ accretes at $f_{\rm Edd} \approx 0.5$, with a slight decrease with time. 
$M_2$, instead, behaves differently. 
In run A, during the first 7 Myrs, on average $f_{\rm Edd}\approx 0.4$.
$M_2$ accretion history can be divided in two phases:\\
1a) for $t\lsim 2.5$ Myrs the circularization process is
still efficient. During this phase $f_{\rm Edd} \approx 0.3$ on average,
showing strong variability;\\   
2a) for $t\gsim 2.5$ Myrs, $M_2$ is
moving on a quasi--circular orbit, and the relative velocity between
$M_2$ and the gaseous disc is reduced. In this phase, 
$f_{\rm Edd} \approx 0.45$ on average.\\ 
Larger variability of $f_{\rm Edd}$ can be observed 
onto the counter--rotating
$M_2$ in run B. We can still distinguish two phases:
1b) for $t\, \lsim \, 3$ Myr, $M_2$ is still counter--rotating
($L_{\rm z} < 0$), and $f_{\rm Edd}\approx 0.1$, well below
$f_{\rm Edd}\approx 0.25$ obtained averaging over 5 Myr;\\
2b) for $t\, \gsim \, 3$ Myr, $M_2$ accretes at  $f_{\rm Edd} \approx 0.45$
(on average).

\section{Conclusions}

Our two high resolution runs show that the gas--dynamical interaction
between the massive circumnuclear disc and  MBHs is unable to
bring the two MBHs at separations of the order of the force resolution
($\approx 0.1$ pc) in $\approx 5$ Myr.
Circularization of the initially eccentric orbit of $M_2$ is efficient until the MBHs form a
binary. In the latest stages of our simulations l
eccentricity grows again up to $\gsim\, 0.1$.
These results are strictly connected to the formation of a central core in the
surface density profile of the circumnuclear disc, due to the energy
and angular momentum exchange between the MBH pair and the gaseous
particles. We stress that we could catch the formation of the central core 
(and the stalling of the binary) because of the high spatial resolution adopted from the beginning of the
simulation, which was increased by a factor $\approx 10$ compared
to our previous simulations. It must be pointed out that, as discussed in Mayer et al.
2007, the fate of the MBH binary could be strongly affected 
by cooling/heating processes, not implemented in our runs.

Thanks to the high spatial resolution of our simulations that allows for 
the code to resolve the sphere of influence of the MBHs, we studied
for the first time the mass accretion rate as a function of the
dynamical properties of the MBHs: we found that variable
double nuclear activity can be observable for few Myr,
when the two MBHs orbit with relative separations $\approx 10$ pc.
The accretion rate on counter--rotating orbits 
is more variable, and lower by a factor of 4-5 when compared 
to $M_2$ co--rotating with the disc.

\bibliographystyle{aa}

\end{document}